\documentstyle[aps,amssymb,twocolumn,psfig]{revtex}

\def\doteqdot{=}

\begin{document}
\flushbottom
\title{Generalized quantum XOR-gate for quantum teleportation and
state purification
in arbitrary dimensional Hilbert spaces}
\author{Gernot Alber$^1$, Aldo Delgado$^1$, Nicolas Gisin$^2$, Igor Jex$^{1,3}$}
\address{
$^1$ Abteilung f\"ur Quantenphysik, Universit\"at Ulm,
D--89069 Ulm, Germany\\
$^2$ Group of Applied Physics, University of Geneva, 1211 Geneva 4, Switzerland\\
$^3$ Department of Physics, FJFI \v CVUT,
B\v rehov\'a 7, 115 19 Praha 1 - Star\'e M\v{e}sto, Czech Republic}
\date{\today}
\maketitle
\begin{abstract}
A generalization of the quantum XOR-gate is presented
which operates in
arbitrary dimensional Hilbert spaces.
Together with one-particle Fourier transforms this gate is capable
of performing a variety of tasks which are
important for quantum information processing in arbitrary dimensional Hilbert
spaces. Among these tasks are the preparation of Bell states,
quantum teleportation
and quantum state purification.
A physical realization of this generalized XOR-gate is
proposed which is based on non-linear optical elements.
\end{abstract}
In quantum information processing the quantum XOR-gate \cite{Monroe}
plays a fundamental
role. In this 2-qubit gate, the first qubit controls the target qubit: if the control is in
state $|0>$, the target is left unchanged, but if the control qubit is in state $|1>$
the target's basis states are flipped. 
Together with one-qubit operations it forms a universal set of quantum
gates allowing the implementation of arbitrary unitary operations acting on qubits
\cite{gates}. 
It has been demonstrated that it can be used for many practical tasks of
quantum information processing with qubits, such as
quantum state swapping \cite{swapping},
entangling quantum states \cite{controled swapping},
performing Bell measurements \cite{bell measurement}, dense coding \cite{dense}
and teleportation \cite{tele}.
Furthermore, in combination with selective
measurements it can be used for implementing non-linear quantum transformations
of quantum states which may be used for optimal state identification and
for state purification \cite{Bennett,gisin}. 

For many practical tasks of quantum information processing it is
desirable to extend the basic notion of such a quantum XOR-operation to
higher dimensional Hilbert spaces. Indeed, most of the physical systems that have been
proposed to hold qubits, such as
multilevel atoms or ions \cite{multilevel}
and multipath-interferometers \cite{Brendel99}, 
could equally well encode larger alphabeths. 
However, there is a considerable
degree of freedom involved in such a generalization. 
Not all such generalizations of the basic quantum XOR-gate of qubits 
acquire a similar fundamental significance in connection with the universality
of quantum operations in higher dimensional Hilbert spaces.
A proper generalization is thus a useful tool as it allows one
to unify various quantum operations which are of current interest for quantum
information processing, such as entangling quantum states,
performing Bell measurements,
teleportation and purifying quantum states.

In this letter a generalized XOR-gate is proposed
which acts on two arbitrary dimensional quantum systems and which
inherits all the significant properties of the basic XOR-gate for
qubits. In particular we demonstrate that
this generalized quantum XOR-gate
may be used to
entangle two quantum systems with one another, to teleport an unknown
quantum state, and to implement
non-linear quantum transformations for state purification.
A possible physical realization of this quantum gate is proposed which
is based on non-linear optical elements.

Let us start by summarizing characteristic properties of the
XOR-gate as they are known for qubit systems.
For qubits the action of the quantum XOR-gate 
onto a chosen set of basis
states $\left\{ \left| i\right\rangle \right\}$ with $i\in \{0,1\}$
of the Hilbert space
of each qubit is defined by
\begin{equation}
XOR_{12}\left| i\right\rangle _{1}\left| j\right\rangle _{2}\doteqdot \left|
i\right\rangle _{1}\left| i\oplus j\right\rangle_{2}~.
\label{quantum xor-gate}
\end{equation}
This transformation has the
following characteristic properties:
(i) it is unitary and thus reversible,
(ii) it is
hermitian and
(iii) $i\oplus j=0$ if and only if $i=j$. 
The first (second) index denotes the state of the control (target) qubit 
and $\oplus $ denotes addition modulo(2).


Let us now consider the problem of generalizing the quantum XOR-gate to
higher dimensional Hilbert spaces.
The desired generalized quantum XOR-gate (GXOR-gate) should
act on two $D$ -dimensional quantum systems.
In analogy with qubits we will call these two systems 
qudits.
The basis states $|i\rangle $ of each qudit are labeled by elements in
the ring ${\sf Z}_{D}$ which we denote by the numbers $i=0,...,D-1$ with the
usual rules for addition and multiplication {\it modulo(D)}. In
principle, the GXOR-gate could be defined in a straightforward way
by using 
Eq.(\ref{quantum xor-gate}) and by performing $i\oplus j$ 
{\it modulo(D)}, i. e.
\begin{equation}
GXOR_{12}\left| i\right\rangle _{1}\left| j\right\rangle _{2}\doteqdot \left|
i\right\rangle _{1}\left| i\oplus j\right\rangle _{2}.
\label{error}
\end{equation}
However, with this GXOR-gate one cannot purify quantum states with the help
of non-linear quantum transformations as 
$-i\neq i$ in ${\sf Z}_{D}$ for $D>2$. Moreover, the GXOR-gate defined in 
(\ref{error}) is unitary but not hermitian for $D>2$. Therefore it is no longer
its own inverse. 
Thus, the inverse GXOR-gate has to be obtained from the 
GXOR-gate of Eq.(\ref{error}) by iteration, i.e.
$GXOR_{12}^{-1}=(GXOR_{12})^{D-1}=GXOR_{12}^{\dagger}\neq GXOR_{12}$.
All these inconvenient properties of this preliminary definition (\ref{error})
can be removed by the alternative definition
\begin{equation}
GXOR_{12}\left| i\right\rangle_{1}\left| j\right\rangle _{2}\doteqdot \left|
i\right\rangle _{1}\left| i\ominus j\right\rangle _{2}.
\label{generalized xor-gate}
\end{equation}
In Eq.(\ref{generalized xor-gate}) $i\ominus j$ denotes the difference
$i-j$ modulo (D). In the special case of qubits the definition of
Eq.(\ref{generalized xor-gate}) reduces to Eq.(\ref {quantum xor-gate})
as $i\ominus j\equiv i\oplus j~modulo(2)$. Furthermore, this 
definition preserves all the properties of Eq.(\ref{quantum xor-gate}) 
also for arbitrary values of $D$, namely it is unitary,
hermitian and $i\ominus j=0 ~modulo(D)$ if and only if $i=j$. 

The GXOR-gate of Eq.(\ref{generalized xor-gate})
admits a natural extension to control and target systems 
with continuous spectra. In this case the basis states $|i\rangle$
are replaced
by the basis states $\left\{ \left| x\right\rangle \right\}$ with
the continuous variable $x \in {\bf R}$.
These new basis states are assumed to satisfy the 
orthogonality condition $\left\langle x \right| y \rangle=
\delta\left( x-y \right)$.
Furthermore, as the dimension $D$ tends to infinity the modulo
operation entering Eq.(\ref{generalized xor-gate}) can be omitted.
Thus, for continuous variables the action of the GXOR-gate becomes
\begin{equation}
GXOR_{12}\left| x\right\rangle_{1}\left| y\right\rangle_{2}=
\left| x\right\rangle_{1}\left| x-y\right\rangle_{2}.
\label{continuous xor-gate}
\end{equation}
Let us note that this definition for the case of continuous variables
is different from the 
generalized XOR-gate proposed in 
Ref. \cite{braunstein}. This latter gate is not
hermitian whereas the GXOR-gate of
Eq.(\ref{continuous xor-gate}) is both unitary and hermitian.
The GXOR-gate of Eq. 
(\ref{continuous xor-gate}) can be represented in terms of
a translation 
and a space inversion, namely
\begin{equation}
GXOR_{12}\left| x\right\rangle_{1}\left| y\right\rangle_{2}=
\hat{\Pi}_{2}
{\it e}^{(-i\hat{P}^{(2)}_{y}\hat{x}^{(1)})}
\left| x\right\rangle_{1}
\left| y\right\rangle_{2}.
\end{equation}
Thereby $\hat{P}^{(2)}_{y}$ denotes the canonical momentum operator
which is conjugate to the
position operator $\hat{y}^{(2)}$ acting on quantum system 2 
and 
$\hat{\Pi}_2$ is the corresponding operator of space inversion.

With the help of the GXOR-gate of Eq.(\ref{generalized xor-gate})
a variety of quantum operations can be implemented which are of
central interest for quantum information processing.
As a first application let us consider the preparation of a 
basis of entangled states from  separable ones. If $|l\rangle |m\rangle$
with $l,m,=0,...,D-1$ denotes an orthonormal basis of separable states 
an associated basis of entangled 
two-particle states is given by
\begin{eqnarray}
|\psi_{lm}\rangle = GXOR_{12}[(F|l\rangle)_1|m\rangle_2].
\label{Bell}
\end{eqnarray}
Thereby $F$ denotes the discrete Fourier transformation, i.e.
$F|l\rangle = (1/\sqrt{D})\sum_{k=0}^{D-1} {\em exp}(i 2\pi lk/D) |k\rangle$.
For qubits this unitary quantum transformation leads to the well known basis
of four Bell states.
In the simplest higher dimensional case of $D=3$,
for example, the first few states of this entangled
generalized Bell basis are given by
\begin{eqnarray}
|\psi_{00}\rangle &=&\frac{1}{\sqrt{3}}[|00\rangle + |11\rangle +|22\rangle],\nonumber\\
|\psi_{10}\rangle &=&\frac{1}{\sqrt{3}}
[|00\rangle + e^{i2\pi/3}|11\rangle +e^{-i2\pi/3}|22\rangle],\nonumber\\
|\psi_{20}\rangle &=&\frac{1}{\sqrt{3}}
[|00\rangle + e^{-i2\pi/3}|11\rangle +e^{i2\pi/3}|22\rangle],\nonumber\\
|\psi_{01}\rangle &=&\frac{1}{\sqrt{3}}[|02\rangle + |10\rangle +|21\rangle],...
\end{eqnarray}
As the GXOR-gate is hermitian it can also be used to disentangle this basis
of generalized Bell states again by inverting Eq.(\ref{Bell}).
This basic disentanglement property is
of practical significance as it enables one to reduce Bell
measurements to measurements of separable states.
Examples where these latter types of
measurements are of central interest are
dense coding \cite{dense} and quantum teleportation schemes
\cite{tele}.

\begin{figure}
\centerline{\psfig{figure=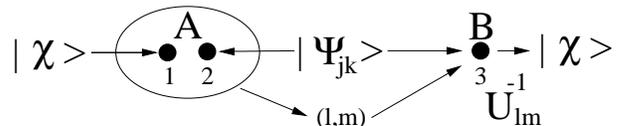,width=8.0cm,clip=}}
\caption{
Schematic representation of the teleportation scheme involving 
Bell measurements 
onto the generalized Bell states of Eq.(\ref{Bell}).
}
\label{Fig.1}
\end{figure}

The basis of entangled Bell
states resulting from Eq.(\ref{Bell}) can be used for teleporting an arbitrary
D-dimensional quantum state from A (Alice) to B (Bob).
For this purpose let us assume that A and B share an entangled pair of
particles prepared in state
$|\psi_{lm}\rangle$ as defined by Eq.(\ref{Bell}). If A wants to teleport an
unknown quantum state $|\chi\rangle = \sum_{n=0}^{D-1}\alpha_n |n\rangle$ to B
she has to perform a Bell measurement which yields one of the entangled basis
states of Eq.(\ref{Bell}) as an output state (compare with Fig. (\ref{Fig.1})). 
Conditioned on the measurement result of Alice, Bob has to perform an appropriate unitary
transformation onto his particle which prepares this latter
particle in state $|\chi\rangle$.
This arbitrary dimensional teleportation scheme rests on the identity
\begin{eqnarray}
|\chi\rangle |\psi_{j k}\rangle_{23} &=&
\sum_{l, m = 0}^{D-1} |\psi_{l m}\rangle_{12}
\frac{e^{-i2\pi j m/D}}{D} U_{l m} |\chi\rangle,\nonumber\\ 
U_{l m }|n\rangle  &=& 
e^{-i2\pi n(l - j)/D} |n - k - m\rangle.
\label{tele}
\end{eqnarray}
This basic relation for teleportation for an arbitrary dimensional state
$|\chi\rangle$ 
can be derived in a straightforward way from Eqs.
(\ref{generalized xor-gate}) and (\ref{Bell}). 
The classical communication requires $2\log_2(D)$ bits, which is the minimum
necessary in all quantum teleportation schemes.

Together with filtering measurements acting on a target quantum system $t$
the GXOR-gate of Eq. (\ref{generalized xor-gate}) can be used to implement
non-linear transformations of quantum states of a control system $c$. This
can be demonstrated most easily by considering the case of two qudits which
are prepared in the quantum states $\sigma^{t}$ and $\sigma^{c}$ initially. 
Let us perform the quantum operation
\begin{eqnarray}
T(\sigma ^{c},\sigma ^{t}) &\doteqdot &\frac{A\left( \sigma ^{c}\otimes
\sigma ^{t}\right) A^{\dagger }}{Tr[A\left( \sigma ^{c}\otimes \sigma
^{t}\right) A^{\dagger }]}  
\label{non-linear transformation 1}
\end{eqnarray}
on these two qudits with
\begin{eqnarray}
A &\doteqdot &({\bf 1_{c}}\otimes P_{-})~GXOR_{ct}. 
\label{Ain}
\end{eqnarray}
Thereby ${\bf 1_{c}}$ denotes the identity operator acting in the Hilbert
space of the control system and $P_{-}=\left|0\right\rangle_{tt}\left
\langle 0\right|$ is the projector onto state $\left| 0\right\rangle_{t}$ of the 
target qudit.
With the
decomposition 
\begin{eqnarray}
\sigma^{c}=\sum_{ij}^{D-1}\sigma_{ij}^{c} 
\left| i\right\rangle_{cc}\left \langle j\right|,\nonumber
\\
\sigma^{t}=\sum_{ij}^{D-1}\sigma_{ij}^{t} 
\left| i\right\rangle_{tt}\left \langle j\right|
\end{eqnarray}
Eqs. (\ref{non-linear transformation 1}) and 
(\ref{Ain}) may be rewritten in the form
\begin{equation}
T(\sigma^{c},\sigma^{t})=\frac{\sum_{ijkl}^{D-1}\sigma_{ij}^{c}\sigma_{kl}^{t}
\left| i\right\rangle_{cc}\left \langle j\right|\otimes
P_{-}\left| i\ominus k\right\rangle_{tt}\left \langle j\ominus l\right|P_{-}}
{\sum_{ikl}^{D-1}\sigma_{ii}^{c}\sigma_{kl}^{t}
\left\langle 0| i\ominus k\right\rangle_{tt}\left \langle i\ominus l|0\right\rangle}.
\end{equation}
Assuming that both control and target qudit are 
prepared in the same state initially, i.e. 
$ \sigma^{c}\equiv \sigma^{t}$,
and using the basic property $i\ominus j=0~{\it modulo(D)}$ if and only if 
$i=j$ of the GXOR-gate of Eq.(\ref{generalized xor-gate}) it turns out that
Eq.(\ref{non-linear transformation 1}) is equivalent to the relations 
\begin{eqnarray}
T(\sigma^{c},\sigma^{t}&\equiv &\sigma^{c})=
\sigma_{output}^{c}\otimes P_{-},\nonumber\\
\sigma_{output}^{c}&=&\frac{\sum_{ij}^{D-1}\left( \sigma_{ij}^{c} \right
)^{2}\left| i\right\rangle_{cc}\left \langle j\right|} {\sum_{i}^{D-1}\left(
\sigma_{ii}^{c} \right )^{2}}.  
\label{final state}
\end{eqnarray}
As a result of the quantum operation (\ref{non-linear transformation 1})
the combined system formed by the control and the target qudit forms a
factorizable state with the target qudit being
in state $|0\rangle \langle 0|$.
According to Eq.(\ref{final state}) the density matrix elements of $\sigma^c$
with respect to the computational basis $|i\rangle$ $(i=0,...,D-1)$ have been
squared. This final state is prepared with probability
$p_c = \sum_i^{D-1}(\sigma_{ii}^c)^2$.
From Eq. (\ref{final state}) it is easy to verify that the quantum 
operation (\ref{non-linear transformation 1}) has the following basic
properties: (i) it maps density matrices onto density matrices, (ii) it is 
not injective and non-linear, (iii) there are states invariant under the 
transformation, and (iv) it maps pure states onto pure states. 
It is also possible to extended the quantum operation of Eq. (\ref
{non-linear transformation 1}) to cases in which there is more than one
control system and in which both the control and the
target systems are composite quantum systems each of which consists of M
qudits. In this case $\sigma ^{c}$ describes a general $M$-qudit state of
the form
\begin{equation}
\sigma^{c}=\sum_{{\bf ij}}\sigma_{{\bf ij}}^{c}\left| {\bf i}%
\right\rangle_{cc}\left \langle {\bf j}\right|,
\end{equation}
with ${\bf i}=(i_1,...,i_M)$ and ${\bf j}=(j_1,...,j_M)$. In Eq.(\ref
{non-linear transformation 1}) the operator $A$ has to be replaced by
\begin{eqnarray}
A &\doteqdot &({\bf 1_{c}}\otimes P_{-})\Pi
_{j=1}^{M}\Pi _{i=1}^{N}GXOR_{ct_{i}}^{(j)}  
\label{finA}
\end{eqnarray}
with the projection operators $P_{-}=\Pi_{i=1}^{N}\otimes
 P_{t_{i}}$ and $P_{t_{i}}=\left| {\bf 0}
\right\rangle_{t_{i}t_{i}}\left\langle {\bf 0}\right|$ onto state $\left| 
{\bf 0}\right\rangle_{t_{i}}$ of the $M$-qudit target system $t_i$. Thereby
the GXOR-gate $GXOR^{(j)}_{ct_i}$ operates on the $j$-th qudit of the control
and of the $i$-th target system.
The resulting final state of the control system is given by
\begin{equation}
\sigma_{output}^{c}=
\frac{\sum_{{\bf ij}}(\sigma_{{\bf ij}}^{c})^{1+N}\left| 
{\bf i}\right\rangle_{cc}\left \langle {\bf j}\right|}
{\sum_{{\bf i}%
}(\sigma_{{\bf ii}}^{c})^{1+N}}.
\label{final register state arbitrary powers}
\end{equation}
and is prepared with probability
$p_c = \sum_{{\bf i}} (\sigma^c_{{\bf ii}})^{1+N}$.

\begin{figure}
\centerline{\psfig{figure=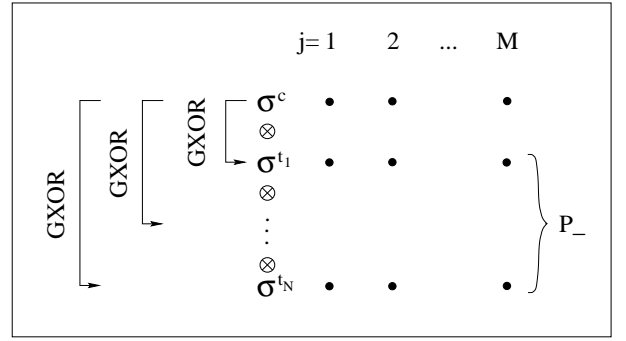,width=8.0cm,clip=}}
\caption{Schematic representation of the GXOR-gates and projections
involved in the non-linear quantum transformation 
of Eq. (\ref{final register state arbitrary powers}). The qudits are represented
by dots. The dots of the first line represent the $M$ qudits of the control
system. The dots of the following lines represent the $M\times N$ qudits of the $N$
target systems $t_{1},t_{2},...,t_{N}$. The GXOR-gate 
$GXOR_{ct_i}^{(j)}$
acts on the $j-th$ qudit
of the control and target system $t_{i}$ with
$j\in\left\{1,2,...,M\right\}$ and $i\in\left\{1,2,...,N\right\}$. The 
operator $P_{-}$ projects the state of the whole
systems onto state $\left| {\bf 0}\right\rangle \left\langle {\bf 0}\right|$ 
with $\left| {\bf 0}\right\rangle=\left| 0\right\rangle_{1}
\left| 0\right\rangle_{2}...\left| 0\right\rangle_{MN}$.}
\label{Fig.2}
\end{figure}

In general, also the non-linear quantum transformation of Eq.
(\ref{final register state arbitrary powers})
has invariant states. This suggests to use this 
non-linear quantum transformation for the purification of quantum states of a
two-qudit system. 
For the special case of a control system consisting of two-qubits
such a purification scheme has already been
proposed previously \cite{gisin}. 
In order to discuss an analogous purification scheme in arbitrary dimensional
Hilbert spaces
we start from the observation that for $M=2$ the entangled basis
state $|\psi_{00}\rangle$ of Eq.(\ref{Bell}) is a fixed point of the non-linear
two-particle quantum map of Eq.(\ref{final register state arbitrary powers}).
Thus this map may be used to purify quantum states towards
the entangled  state $|\psi_{00}\rangle$.
In order to exemplify the convergence properties of this purification process
let us assume that initially we start from a Werner state of the form
\begin{equation}
\sigma^{c} = \lambda |\psi_{00}\rangle \langle \psi_{00}|
+ (1-\lambda){\bf 1}/D^2.
\label{Werner}
\end{equation}
This state may result from a physical situation where two spatially separated
parties, say A(lice) and B(ob),
want to share the entangled basis state $|\psi_{00}\rangle$
but with a probability of $(1-\lambda)$ the
transmission of this entangled pair leads to unwanted noise represented
by the chaotic state ${\bf 1}/D^2$.
This quantum state $\sigma^{c}$ is non-separable
if and only if $\lambda >\lambda_D = (1 + D)^{-1}$ \cite{Ruben}
so that a purification scheme based on
Eq.(\ref{final register state arbitrary powers}) can succeed only for these
values of
$\lambda$.
In order to maximize the range of
convergence of a purification scheme
based on
Eq.(\ref{final register state arbitrary powers}) let us
introduce an additional unitary twirling transformation
\cite{Horodecki96} of the form
$U_A\otimes U_B^*$
which is performed by parties A and B
locally after each iteration of the non-linear
quantum map (\ref{final register state arbitrary powers}).
Thus, at each step of the purification process the mapping
\begin{equation}
\sigma^{c} \to
U_A\otimes U_B^*
\frac{\sum_{{\bf ij}}(\sigma_{{\bf ij}}^{c})^{2}\left| 
{\bf i}\right\rangle_{cc}\left \langle {\bf j}\right|}
{\sum_{{\bf i}%
}(\sigma_{{\bf ii}}^{c})^{2}}
U_A^{\dagger}\otimes U_B^{*\dagger}
\label{pur}
\end{equation}
is performed. 
Thereby the local unitary transformation redistributes all states.
The only state which is left invariant by this redistribution
procedure is the entangled state $|\psi_{00}\rangle$.
In principle,
the local unitary transformation $U$ can be chosen arbitrarily.
However, numerical simulations indicate that the region of convergence
of the purification process can be improved considerably
by choosing two different types of local
unitary twirling transformations which are used
alternatively.
As an example, let us choose 
for $U$ alternatively a discrete
Fourier transform involving all  $D$ states $|0\rangle, ...|D-1\rangle$
and a discrete Fourier transform involving the $D-1$ states
$|0\rangle, ...|D-2\rangle$ only.
Numerical simulations which we have performed 
for dimensions $2\leq D \leq 20$ demonstrate clearly
that the purification
procedure involving these two local unitary transformations
is capable of purifying all non-separable
Werner states of the form of Eq.(\ref{Werner}).
Thus, it is expected
that this maximal range of
convergence of this purification scheme also applies to all higher
dimensional Hilbert spaces.

Let us finally discuss a possible
physical realization of the 
GXOR-gate defined by Eq.(\ref{generalized xor-gate}) 
which is based on
non-linear optical elements.
For this purpose we assume that the two quantum systems which are going to be
entangled are two modes of the radiation field. The basis states $|i\rangle_1$ $(i=0,...,D-1)$
of the first quantum system are formed by $n$-photon states 
of  mode one with $0\leq n\leq D-1$.
The basis states of the second quantum system $|k\rangle_2$
$(k=0,...,D-1)$ are formed by
Fourier transformed $n$-photon states of this latter mode,
i.e. $|k\rangle_2 = 1/\sqrt{D}\sum_{n=0}^{D-1}
{\rm exp}(i2\pi kn/D)|n\rangle_2$.
Let us further assume that the dynamics of these two modes of the
electromagnetic field are governed by the Kerr-effect \cite{Kerr}. Thus, in the interaction 
picture their Hamiltonian  is given by
$H=\hbar\chi a_1^{\dagger}a_1 a_2^{\dagger}a_2$ with the creation and annihilation
operators $a_{1,2}^{\dagger}$ and $a_{1,2}$ of modes $1$ and $2$, respectively.
For the sake of simplicity 
the nonlinear susceptibility $\chi$ is assumed to be real-valued and positive.
Preparing intitially both quantum systems in state
$|i\rangle_1|k\rangle_2$ after an interaction time
of magnitude $t=2\pi/(D\chi)$ 
this two-mode system ends up in state
$|\psi\rangle_{12} = |i\rangle_1|k-i\rangle_2$. Applying to this latter state
a time reversal transformation which may be implemented by the process of
phase conjugation \cite{Kerr}
we finally arrive at the desired state $|i\rangle_1|i-k\rangle_2$.
Thus this combination of a Kerr-interaction with a time reversal transformation
is capable of realizing the GXOR-gate of Eq. (\ref{generalized xor-gate}).

In summary,
a generalized quantum XOR-gate has been proposed 
which acts on two quantum systems in arbitrary dimensional Hilbert spaces.
This quantum gate is
unitary and hermitian 
and preserves characteristic properties of the basic quantum XOR-gate
acting on qubits.
It has been demonstrated that together with one-particle Fourier-transformations
this GXOR-gate
is capable of performing various important elementary tasks of quantum information
processing in arbitrary dimensional Hilbert spaces, such as
the preparation of entangled basis states (the so-called Bell states),
quantum teleportation and quantum
state purification. Physically
this proposed quantum gate can be implemented  optically, for example,
with the help of the Kerr-effect and with the help of phase conjugation.
These applications 
demonstrate the usefulness of the presented GXOR-gate
as a basic and unifying concept for several problems of quantum information
processing in arbitrary dimensional Hilbert spaces.

This work is supported by
the DFG (SPP `Quanteninformationsverarbeitung'),
by the ESF programme on
`Quantum Information Theory and Quantum Computation' and by the
European IST-1999-11053 EQUIP project.
A.D. acknowledges support by the DAAD.
I.J is supported by the A. von Humboldt foundation and 
by the Ministry of Education of the Czech Republic. 
We are grateful to Steve Barnett for stimulating discussions.


\begin{references}
\bibitem{Monroe}  C.~Monroe et al., Phys. Rev. Lett. {\bf 75}, 4714 (1995).
\bibitem{gates}  A.~Barenco et al.,
Phys. Rev.
Lett. {\bf 74},  4083 (1994); D.~ DiVincenzo, Phys. Rev. A {\bf 51}
 1015 (1995); A.~Barenco et al.,
Phys. Rev. A {\bf 52},  3457 (1995).
\bibitem{swapping}  R.~Feynman, Opt. News {\bf 11},  11 (1985).
\bibitem{controled swapping}  A.~Barenco et al.,
SIAM Journal of Computing {\bf 26},
 1541 (1997).
\bibitem{bell measurement}  S.~L.~Braunstein, A.~Mann, M.~Revzen, Phys. Rev.
Lett. {\bf 68},  3259 (1992).
\bibitem{dense}  C.~H.~Bennett and S.~J.~Wiesner, Phys. Rev. Lett. {\bf 69},
 2881 (1992).
\bibitem{tele} C.~H.~Bennett et al.,
Phys. Rev. Lett. {\bf 70},  1895 (1993). 
\bibitem{Bennett}  C.~ H.~ Bennett et al.,
 Phys. Rev. Lett. {\bf 76}, 722 (1996).
\bibitem{gisin}  H,~Bechmann-Pasquinucci, B.~Huttner, N.~Gisin, Phys. Lett.
A {\bf 242}, 198 (1998).
\bibitem{multilevel} Ch.~Roos et al.,
Phys. Rev. Lett. {\bf 83}, 4713 (1999). 
\bibitem{Brendel99}J. Brendel, N. Gisin, W. Tittel, and H. Zbinden. 
        Phys. Rev. Lett. {\bf82}, 2594 (1999).
\bibitem{braunstein} S.~L.~Braunstein,
Phys. Rev. Lett. {\bf 80}, 4084 (1998).
\bibitem{Ruben} A. O. Pittenger and M. H. Ruben, quant-ph/0001110.
\bibitem{Horodecki96} M.~ Horodecki and P.~ Horodecki,
Phys. Rev. A {\bf 59}, 4206 (1999). 
\bibitem{Kerr} L. Mandel and E. Wolf, {\em Optical Coherence and Quantum Optics},
(Cambridge, Cambridge, 1995).
\end{references}
\end{document}